\documentstyle[multicol,epsfig,prl,aps]{revtex}
\tighten
\draft

\begin{document}
\title{Electronic correlations in transport through coupled quantum dots} 
 
\author{Antoine Georges$^1$, Yigal Meir$^2$} 
\address{$^1$Laboratoire de Physique Th{\'e}orique de l'Ecole 
Normale Sup{\'e}rieure,\cite{auth1}
24 rue Lhomond, 75231 Paris Cedex 05, France}
\address{$^2$ Physics Department, Ben Gurion University, Beer Sheva 84105, 
Israel}
\date{\today}
\maketitle
\begin{abstract}
The conductance through two quantum dots in series is studied using general
qualitative arguments and quantitative slave-boson mean-field theory. It is
demonstrated that measurements of the conductance can explore the phase
diagram of the two-impurity Anderson model. Competition between the Kondo
effect and the inter-dot magnetic exchange leads to a two-plateau structure
in the conductance as a function of gate voltage and a two or three 
peak structure in the conductance vs. inter-dot tunneling.
\end{abstract}

\def\eL{\epsilon_{0L}}
\def\eR{\epsilon_{0R}}
\let\e=\epsilon
\def\Im{\mathop{\hbox{Im}}}
\def\Re{\mathop{\hbox{Re}}}
\def\xj{J/T_K^0}
\def\xt{t/\Gamma}
\def\Ee{\overline{E}_e}
\def\Eo{\overline{E}_o}
\def\Ea{\overline{E}_{\alpha}}
\def\de{\delta_e}
\def\do{\delta_o}
\def\squote{}
\def\quote#1#2#3#4{\squote {#1,\ {\sl#2}\ {\bf#3}, #4}.\par} 
\def\qquote#1#2#3#4{\squote {#1,\ {\sl#2}\ {\bf#3}, #4};} 
\def\nquote#1#2#3#4{\squote {#1,\ {\sl#2}\ {\bf#3}, #4}} 
\def\book#1#2#3{\squote { #1,\ in {\sl#2}, edited by #3}.\par}
\def\nbook#1#2#3{\squote { #1,\ in {\sl#2}, edited by #3}}
\def\bbook#1#2#3{\squote { #1,\ in {\sl#2}, edited by #3}.}
\def\trans#1#2#3{[ {\sl #1} {\bf #2},\ #3 ]}
\def\prl{{\sl Phys. Rev. Lett.}\ }
\def\PRL{{\sl Phys. Rev. Lett.}\ }
\def\apl{{\sl Appl. Phys. Lett.}\ }
\def\APL{{\sl Appl. Phys. Lett.}\ }
\def\euro{{\sl Europhys. Lett.}\ }
\def\ds{\displaystyle}
\def\pr {{\sl Phys. Rev.}\ }
\def\ksi{\xi}
\def\l{\ell}
\def\sss{\scriptscriptstyle}
\def\w{\omega}
\def\e{\epsilon}
\def\deriv{\partial}

\begin{multicols}{2}

The recent observation of the Kondo effect in transport through a quantum dot
\cite{goldhaber,kouwenhoven,reviews} 
opened a new path for the investigations of strongly correlated
electrons. Having confirmed earlier theoretical predictions 
\cite{glazman88,hershfield91},  that a quantum dot behaves as
a magnetic impurity,  these experiments also serve as a critical quantitative
test for existing theories. In particular,  unlike magnetic impurities in metals
which have physical properties determined by the host metal and the impurity 
atom, the corresponding parameters in the quantum dot case can be varied
continuously,  enabling,  for example, a crossover from the Kondo to the
mixed-valence and the empty dot regimes in the same sample 
\cite{goldhaber,kouwenhoven}. 

The behavior of a lattice of magnetic impurities, such
as a heavy-fermion system, is characterized by the competition
between the Kondo effect and the magnetic correlations between
the impurities. An important step towards the understanding of this problem
 was taken by Jones and collaborators \cite{jones},  who studied the 
 two-impurity problem.
Their work demonstrated that this competition 
leads to a second-order phase 
transition when particle-hole symmetry applies. When this symmetry is broken, 
this transition is replaced by a crossover \cite{MKJ,SaSh,ALJ}. In view of
the extensive experimental research on transport through two dots in series
 \cite{twwdotsexp,twwdotsrecent},  
it is thus natural to try and understand how this
phase-transition is manifested in the  double-dot system, 
both because such systems may have important applications (such as a quantum-dot
laser \cite{nedlaser}), and because such a tunable system may reveal detailed
information on the corresponding phase diagram. 

\begin{figure}[]
\epsfig{file=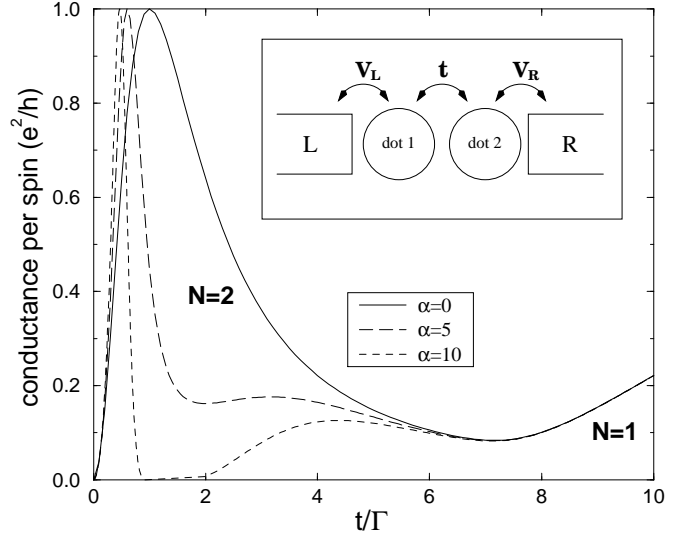,width=10cm,angle=0}
\caption[0]{\narrowtext Plot of the conductance vs. the tunneling between the 
dots, $t$,  obtained by slave-boson mean-field theory.
 Due to the Kondo effect in the two-electron regime ($N=2$)
 the conductance has a peak at $t=\Gamma$. 
As $t$ increases beyond $\overline{t_{+}}$, the Kondo effect is quenched, 
 until the ground-state contains a single electron ($N=1$), 
leading to a different
 Kondo state and an enhanced conductance. 
 For finite antiferromagnetic coupling $J$ (finite $\alpha=\Gamma^2/UT_K^0$),
 the conductance 
peak is pushed to smaller values of $t$ and becomes narrower,  
as the singlet formation destroys the Kondo state. In
 addition a second maximum in the conductance in the $N=2$ regime emerges.
 {\sl Inset}: The double dot system discussed in this paper.}
\label{conductancet}
\end{figure}

Transport through a double-dot system (see inset in Fig.~\ref{conductancet})
has already received 
much theoretical attention,  in particular in the high temperature,  Coulomb
blockaded regime\cite{nedlaser,2dottheory}. In experiments
the Coulomb charging energy and the excitation energy are much larger than
temperature. Accordingly, only a single state on each dot is important,
and double occupancy of each dot can be ignored.
Denoting the energies of these states by $\e_1\equiv\e_0+\Delta\e$ 
and $\e_2\equiv\e_0-\Delta\e$, respectively, and the tunneling amplitude
between the dots by $t$,
the isolated double dot system can contain zero, one or two electrons, 
depending on
the chemical potential: $N=0$ for $\mu<\e_-$, $N=1$ for $\e_-<\mu<\e_+$,
and $N=2$ for $\mu>\e_+$, with $\e _{\pm}=\e_0\pm\sqrt{\Delta\e ^2+t^2}$.
In the presence of a  
finite antiferromagnetic spin-exchange $J$ between the dots,  one still has 
 the above three possibilities with
$\e_+$ replaced by $\e_+-3J/4$ \cite{largeJ}.
In the Coulomb blockade regime there will be two peaks in
the conductance vs. chemical potential at the degeneracy points $\mu=\e _{\pm}$.
Alternatively, starting from the $N=2$ regime for $t=0$, $\e_+$ will increase
with increasing $t$, until it crosses the chemical potential and the ground-state
will have a single electron in the double dot system. Again we expect a peak in
the conductance at $t=t_{+}$ corresponding to $\e_+(t_{+})=\mu$.

At low temperatures,  in addition to a renormalization of the above
 energy scales, $\e_\pm \rightarrow \overline{\e}_\pm$,  and 
 $t_{+} \rightarrow \overline{t_{+}}$,  
 the Kondo effect starts to play a significant role in the
transport.
The relevance of the Kondo effect in the double dot system
has been studied in \cite{ivanov,AEK}. 
Here, we focus on the competition between the Kondo effect and 
antiferromagnetic exchange, and present detailed predictions for the 
conductance.   
(Recently Andrei et al. \cite{andrei}
have investigated a very different realization of this competition,   
which applies to coupled {\it metallic islands},
 close to points of charge degeneracy -- i.e. near the Coulomb-blockade peaks
 \cite{matveev}). 
We find a rich phase diagram
leading to interesting features in the conductance as a function of 
gate voltage and intra-dot tunneling. As the corresponding energy and 
temperature
scales are experimentally accessible,  these predictions are  relevant to
transport experiments in double-dot systems.

To simplify notations, 
we assume in the following $\e_1=\e_2=\e_0$ and $V_L=V_R=V$,  where $V_L (V_R)$
is the coupling to the left (right) lead. 
Then the eigenstates
of the double dot system are the even and odd states. As even-odd symmetry is
broken anyway by the tunneling $t$,  one can show that the above assumptions
have little effect on the underlying physics.
Following Ref.\cite{MW}, the zero-temperature conductance 
{\sl  per spin channel} through the system can be expressed in terms of 
the retarded Green functions for the even and odd states 
${\cal G}^{ret}_{e,o}(\w)$ as: 
$g = {\ds e^2\over h} \Gamma^2 
|{\cal G}^{ret}_e(\omega=0)-{\cal G}^{ret}_o(\omega=0)|^2$  
 where $\Gamma\equiv \pi\rho V^2$
  and $\rho$ is the density of states in the leads at the Fermi
  energy. Defining the corresponding {\it scattering 
 phase-shifts}, 
 $\delta_\alpha\equiv \pi + {\rm arg}{\cal G}^{ret}_\alpha(\mu)$ (such
 that $\delta_\alpha$ is in the $0$-$\pi$ range),  the conductance
 formula simplifies to
\begin{equation}
 g = {\ds e^2\over h}\,\sin^2(\delta_e-\delta_o)
\end{equation}
The Friedel sum rule relates the 
total charge $q$ {\it per spin channel}, on the double dot system, 
to these phase shifts:
$q\,=\,{(\delta_e+\delta_o)/{\pi}}$.
There is,  however,  no individual relation between  
$\delta_{\alpha}$ and the occupation of the corresponding state.

For $\mu>\overline{\e}_{+}$ (or alternatively $t<\overline{t_{+}}$), there
are $N=2$ electrons in the system ($q=1$), and states with $N=0$ or $N=1$ are 
high-energy states that can be eliminated from the Hilbert space. The
effective low-energy Hamiltonian only involves spin degrees
of freedom on the dots. It can be cast into the form of a model of
{\it two Kondo impurities}, with Kondo couplings to the even and 
 odd combinations of the conduction electrons in the leads,
an inter-impurity magnetic exchange $J\propto t^2/U$, and a 
potential scattering term in the leads $V^{e,o}$ such that 
$V_e-V_o\propto t\,V^2/(\epsilon_0^2-t^2)$. 

Previous studies of the two-impurity 
Kondo problem (mainly using Wilson numerical renormalization group (NRG))
 \cite{jones,SaSh} have already yielded information on how the phase shifts
  $\de$, $\do$ 
behave as a function of the couplings at $T=0$  in this regime. 
Let us start
with the case $J=0$. For $t=0$ there is an even-odd symmetry, 
and each channel has its own Kondo effect,  leading to $\de=\do=\pi/2$, 
and naturally a zero conductance.
 As  $t/\Gamma$ is increased, the difference $\de-\do$ increases to reach 
the value $\pi/4$, at which point the conductance takes its maximum 
possible value: $e^2/h$ per spin channel. As $t/\Gamma$ is further 
increased, the Kondo effect is gradually overcome by potential
scattering and one reaches $\do\simeq 0$, $\de\simeq\pi$ in the large 
$t/\Gamma$ limit (but still with $t\ll \overline{t_{+}}$). 
A slave-boson mean-field
theory (SBMFT) presented below (cf. also Ref.\cite{AEK}) 
yields in this regime: 
($J=0$, $q\simeq 1$) $\delta\equiv\delta_e-\delta_o=2\tan^{-1}{t/\Gamma}$, 
leading to 
\begin{equation}
g = {e^2\over h}\,{{4t^2\Gamma^2}\over{(t^2+\Gamma^2)^2}}
\label{zeroJ}
\end{equation}
which reaches its maximum value at $t=\Gamma$ (solid curve 
in Fig.~\ref{conductancet}).
The Kondo temperature 
in this regime is of order 
$T_K=c_1\,T_K^0 e^{c_2 t/\Gamma}$, where the $c$'s 
are weakly dependent on $t/\Gamma$ and $T_K^0 \equiv W 
e^{-\pi|\epsilon_0|/\Gamma}$ is the single-dot Kondo temperature. 
(The SBMFT yields $c_1=\cos\delta/2$ and 
$c_2=\delta/2$). The crucial content of that formula is that the 
coupled-dot Kondo temperature can be {\it much larger than the 
single-dot Kondo temperature} for small $J$ and large 
$t/\Gamma$. This has important consequences for the observability 
of the effects described in this paper.

\begin{figure}[]
\epsfig{file=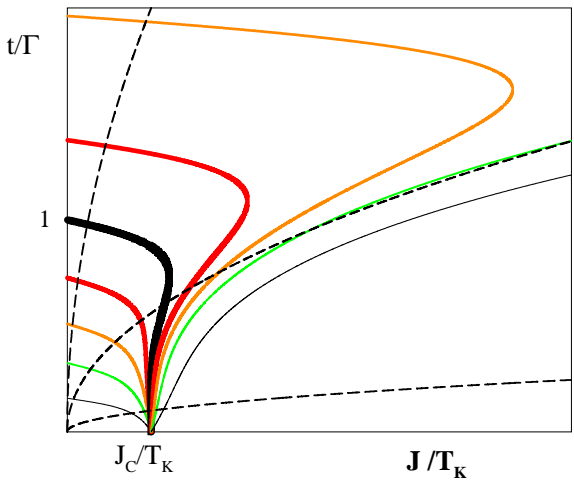,width=9cm,angle=0}
\caption[0]{\narrowtext Schematic contour map of the conductance in 
the $N\simeq 2$ regime. Thicker lines denote higher conductance, 
the thickest one corresponding to $g=e^2/h$. The broken lines are
three physical
contours (for different values of $\alpha\equiv\Gamma^2/UT_K^0$) 
along which $J\sim t^2/U$.}
\label{contour}
\end{figure}

Let us now consider the effect of a finite $J$. For $t=0$. 
 the effective two-impurity Kondo model has particle-hole 
symmetry, and it is known from the work of \cite{jones} that a 
{\it phase transition} exists at a critical value of the coupling  
$J_c/T_K^0\simeq 2.2$. For $J<J_c$, the spin of each dot undergoes a 
Kondo effect with the leads and $\de=\do=\pi/2$. 
For $J>J_c$, the two spins are locked into a singlet state and the 
Kondo effect does not apply, yielding $\delta_o=0, \delta_e=\pi$. 
The phase-shift difference $\delta$ jumps discontinuously from 
$\delta=0$ for $J<J_c$ to $\delta=\pi$ for $J>J_c$. (The conductance is, 
 of course, 
zero for all $J$ since $t=0$). Turning on a small value of $\xt$ is known 
to be a ``relevant perturbation'' on this critical point (with dimension 
$1/2$, identical to that of $J-J_c$) \cite{MKJ,SaSh} and 
therefore smears the transition into a rapid crossover from 
$\delta=0$ to $\delta=\pi$. For $J$ close to $J_c$, this smearing is 
described by a crossover scaling function:
\begin{equation}
{\delta\over\pi}\,= \, \phi\left({{(J-J_c)/T_K^0}\over{t/\Gamma}}\right)
\label{crossover}
\end{equation}
with $\phi(x\rightarrow -\infty)=0$ and $\phi(x\rightarrow +\infty)=1$. 
As a result, the conductance has a very sharp maximum 
as $t/\Gamma$ is increased for a fixed value of $J$ close to $J_c$.  
For $J$ significantly larger than $J_c$, the conductance remains 
very small with only a shallow maximum as $t/\Gamma$ is increased. 
For intermediate values of $t/\Gamma$ and $J/T_K^0$, a quantitative 
calculation of $\delta$ is needed in order to obtain the conductance, 
using e.g. NRG \cite{jones,SaSh,ISS} or SBMFT. However, 
much can be said on a semi-quantitative 
level by using existing knowledge on the two-impurity Kondo problem. 
The phase shift $\delta$ is an increasing function of $J$, which 
starts at the value given above Eq.(\ref{zeroJ}), 
and increases until it saturates at $\delta=\pi$  at 
a scale $J^*$. From the above 
estimate of $T_K$ the ratio $J^*/T_K^0$ increases exponentially with   
$t/\Gamma$. These considerations, and the knowledge of the 
crossover around $J_c$ (Eq.(\ref{crossover})), lead to 
a qualitative contour plot of the conductance in the 
$(J/T_K^0,t/\Gamma)$ parameter space,
throughout the $N\simeq 2$ regime, as displayed in Fig.~\ref{contour}. 

In practice, the exchange $J$ is not an independent parameter, but is a 
function of the interdot tunneling, $J\sim t^2/U$. The contour 
plot above must thus be intersected by a curve 
$J/T_K^0=\alpha (t/\Gamma)^2 $, with $\alpha\equiv\Gamma^2/UT_K^0$, 
in order to follow the dependence of 
the conductance as a function of $t/\Gamma$.
Since $T_K^0$ is a very sensitive function of the energy scales (such as
$\e_0$ and $\Gamma$),  the control parameter $\alpha$  can be varied continuously 
over many orders of magnitude, allowing an experimental investigation of
most of the phase-diagram.
 Thus,  as a function of $t$,  the maximum  conductance $e^2/h$  
is reached for $t\simeq\Gamma$ with a peak width $\Delta t\propto\Gamma$ 
for small $\alpha$, while the peak is pushed down to 
much lower transmission $t\simeq\Gamma/\sqrt{\alpha}$ and becomes very narrow  
$\Delta t\simeq \Gamma/\alpha$ for large $\alpha$.
In addition,  as the saturation scale $J^*$ increases 
exponentially with $t$, one may expect, for an intermediate $\alpha$ (middle
broken curve in Fig.~\ref{contour}), 
 an additional peak in the  conductance vs. $t$ in the $N=2$ regime.
  These results
are indeed confirmed by the SBMFT calculation (see Fig.~\ref{conductancet}).

As $t$ is further increased ($t>\overline{t_{+}}$),   
the equilibrium charge decreases to $N=1$ ($q=1/2$). In
this regime the effective Hamiltonian is that of a 
{\it single-impurity} Kondo problem in the even parity sector \cite{ivanov}, 
leading to unitary scattering $\de\simeq\pi/2$. In the odd parity 
sector, we have an almost empty resonant level with $\delta_o\simeq 0$
(Note that $(\de+\do)/\pi=q \simeq 1/2$ consistent with Friedel sum rule). 
Throughout this regime, we therefore expect the zero-temperature
 conductance to 
be maximum $g = e^2/h$ and essentially independent of $t$. In this 
regime, the inter-dot exchange $J$ plays little role.

\begin{figure}[]
\epsfig{file=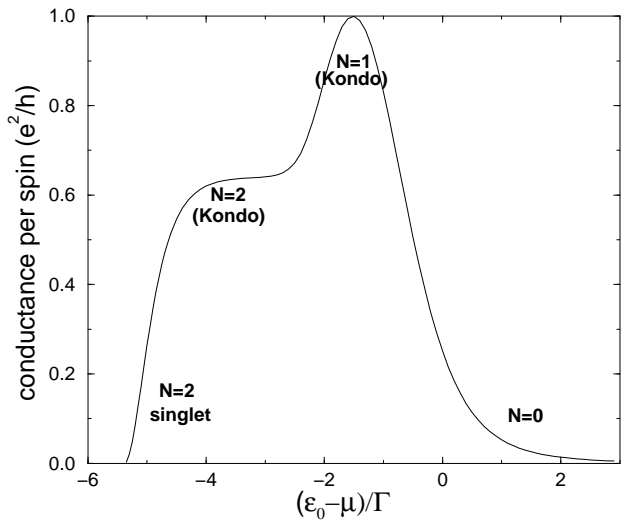,width=9.5cm,angle=0}
\caption[0]{\narrowtext Plot of the conductance vs. the level energy,
 as obtained
from SBMFT (for $t=2$ and $U/\Gamma=10^4$). 
The conductance rises from a very small value (the singlet regime,
$J\gg T_K$), to a $t$-dependent value ($N=2$ Kondo regime, $J\ll T_K$),
 and then to
$g=e^2/h$ ($N=1$ Kondo regime) before dropping to zero for an empty dot.}
\label{conductancegate}
\end{figure}

Similar interesting behavior is expected as a function of gate voltage, 
that controls the depth of the level energy $\e_0$ with respect to the chemical
 potential (see Fig.~\ref{conductancegate}).
  For a very deep level the Kondo temperature is exponentially
 small,  and thus $J/T_K^{(0)}$ is large and quenches the Kondo effect. As $\e_0$
 increases, the  Kondo temperature increases and we enter the ($N=2$) Kondo 
state, and a finite conductance. This conductance remains constant
  (at zero temperature) at a value  {\it smaller than $e^2/h$},
determined by the value of $t$,  until 
  $\e_0$ crosses the Fermi energy and a new ($N=1$) Kondo state is formed.
 There the conductance is given by its maximum value,  $e^2/h$ per spin.
 As  $\e_0$ is further increased the double dot system becomes empty and the 
 conductance drops to zero.

To substantiate these semi-quantitative arguments we have 
performed a quantitative calculation of the phase shifts and
the conductance using a slave-boson mean-field approximation. 
This method becomes exact as the number of spin-degrees of freedom
goes to infinity, and has been 
previously used in order to study the two-impurity Anderson 
model in Ref.\cite{MKJ}. It was recently applied in the present context in
Ref.\cite{AEK} but only in the case $J=0$.
We have solved numerically the full set of equations including 
the tunneling $t$ and exchange $J$, but 
we only quote here the simplified version of the equations \cite{MKJ}
that hold in the $q=1$ regime (one electron in the double dot system 
per spin-state, corresponding to $N=2$).  
For small enough values of $J/T_K^0$, the phase-shift 
difference $\delta$ is given by the solution of:
\begin{equation}
{{2\pi}\over{\delta}} e^{{{\delta t}/{2\Gamma}}} \left(
\sin{\delta\over 2} -{t\over\Gamma}\cos{\delta\over 2}\right)
= {{J}\over{T_K^0}}
\end{equation}
As $J$ is increased beyond a critical coupling $J_c^{SB}$,
$\delta$ reaches the value $\pi$: this is
either a smooth transition for $t>1/\pi$, or a first-order jump for
$t<1/\pi$ (determined by free-energy considerations). The existence
of a phase transition even for non-zero values of $t/\Gamma$ is an
artifact of the SBMFT approximation: $J_c^{SB}$ should actually be
interpreted as an estimate of the saturation scale $J^*$ discussed
above. This spurious transition does not affect qualitatively the
behavior of the conductance, except when it becomes
very small: there a strictly zero-value of $g$ can be found (as 
evident on Fig.~\ref{conductancet}) whereas
the real system would have only a very small but finite $g$.
The SBMFT also provides a quantitative estimate of the Kondo scale for
the coupled dot system in the $q\simeq 1$ regime, as mentioned after
Eq.~(\ref{zeroJ}).

In conclusion, we have demonstrated that measurements of the conductance
through a double-dot system can explore the phase diagram of the
two-impurity Anderson model. By changing the control parameter 
$\alpha=\Gamma^2/UT_K^0$ (which depends sensitively on gate-voltage),
one can make various cuts through the phase-diagram (Fig.~\ref{contour}),
leading to non-trivial features in the conductance vs. gate-voltage and
inter-dot tunneling (Figs.~\ref{conductancet} and \ref{conductancegate}).
As the relevant temperature scale can be much
higher than the single-dot Kondo temperature we believe that these predictions
could be tested experimentally. 



This work was supported in part by a grant from the French-Israeli 
Scientific and Technical Cooperation Program (Arc en Ciel-Keshet-1998). Work
at BGU was further supported by the The Israel Science Foundation - 
Centers of Excellence Program.

\end{multicols}

\begin{references}
\bibitem{largeJ}
 However,  for $J>8\sqrt{\Delta\e ^2+t^2}/3$
there will be a direct jump from $N=0$ to $N=2$. 
Since in practice $J$ stems from superexchange between the dots, $J\sim t^2/U$, 
the latter case may not be experimentally relevant, 
as it requires $t\sim U$.
\end{references}
\end{document}